\documentstyle[prc,aps,eqsecnum,epsfig]{revtex}
\draft
\tightenlines
\def\be {\begin{equation}}
\def\ee {\end{equation}}
\def\bea {\begin{eqnarray}}
\def\eea {\end{eqnarray}}

\begin{document}
\title{
\vspace*{-1cm}
\rightline{\vbox{\hbox{\rightline{LBNL-45199}} 
\hbox{\rightline{McGill/00-12}}}} \vspace*{1cm}
 Baryonic contributions to the dilepton spectra in 
relativistic heavy ion collisions}
\author{M. Bleicher$^*$, A. K. Dutt-mazumder$^\dagger$ , 
C. Gale$^\dagger$, C. M. 
Ko$^\ddagger$ and V. Koch$^*$}
\address{$^*$ Nuclear Science Division, Lawrence Berkeley 
National Laboratory\\ Berkeley, CA 94720, USA\\
$^\dagger$Physics Department, McGill University, Montreal, Quebec
H3A 2T8, Canada\\
$^\ddagger$Cyclotron Institute and Physics Department, 
Texas A$\&$M University\\ College Station, TX 77843, USA}

\maketitle
\begin{abstract}
We investigate the baryonic contributions
to the dilepton yield in high energy heavy ion collisions within the
context of a transport model. The relative contribution of the
baryonic and mesonic sources are examined. It is observed that
most dominant among the baryonic channels is the  decay of $N^\ast(1520)$
and mostly confined in the region below the $\rho$ peak. In a transport
theory implementation we find the baryonic contribution to the lepton
pair yield to be small. 
\end{abstract}
\section {Introduction} \label{sec:1}
One of the main goals of relativistic heavy ion physics is to 
explore the
possibility of studying strongly interacting matter in a highly exotic
form: the quark
gluon plasma. In this context the production of dileptons
in high energy heavy ion collisions has
received particular attention in recent years. This interest is due to
the fact that dileptons, once produced, 
essentially decouple from the strongly interacting system and will 
reach the detectors mostly unscathed. Hence the dileptons serve as
a promising probe, as their production rate is biased towards regions of
high densities and high temperatures. Information on those is only
available indirectly if one is limited to hadronic observables.

It is possible to roughly separate the dilepton invariant mass region in
three parts: the ``soft'' dileptons essentially lie below the $\phi$
peak and the ``hard'' dileptons lie beyond the $J/\psi$. The region 
in-between has been coined the ``intermediate mass region''. It is
important to realize that different physics can be at work in those 
different regimes, mainly reflecting the different epochs of the
relativistic nuclear collisions. For example, the hard dileptons will
receive an important contribution from the Drell-Yan process which will
happen in the first instants of the interaction. In this work, we shall
restrict our attention to the soft part. This region is of interest as
it has been the focus of detailed recent experimental measurements.
Those were done by the HELIOS/3 \cite{helios} and CERES \cite{ceres96} 
collaborations, respectively. Those heavy ion experiments have 
reported an excess
of lepton pairs over measurements involving proton-nucleus collisions.
This has stimulated a great deal of theoretical activity 
\cite{theo,rapp96,song_koch} which has been summarized recently 
\cite{korev,rwreview}.
 
The theoretical interpretation of the dilepton excess has mostly
  concentrated on possible in-medium effects. One school of thought
attributes the low mass abundance to a dropping
vector meson mass, precursor of a chiral symmetry restoration in the
dense medium \cite{br}. Another evaluates the in-medium $\rho$ 
spectral function
and finds it considerably broadened (its peak value is not shifted
appreciably), owing mainly to the coupling of the $\rho$ with baryonic
resonances \cite{friman_pirner}, with the N$^*(1520)$ playing a
dominant role as first pointed out in \cite{lenske}. To be 
precise, the effect is 
distributed over several channels \cite{rapp96,rwreview}. However, the current
  experimental data are also consistent with a scenario without in-medium
  modifications \cite{koch_bormio,koch_hirschegg}. Certainly, data with better
  statistics and an improved mass resolution are needed to extract any
  possible in-medium effects as well as to distinguish between 
the different in-medium modifications.

Several issues deserve further investigations in this matter. The
dilepton calculations based on in-medium spectral functions 
receive a substantial
contribution from the baryonic sector \cite{rapp96}. This feature 
is at first view 
puzzling, since the baryons are a minor component of the hadron
population at CERN energies \cite{js96}. 
A rough estimate of the contribution of baryons to the dilepton
  channel had been provided in \cite{song_koch}, and turned out 
to be about a factor
  of two below the contribution of the $\omega$-Dalitz decay.
With this in mind, we have set
out to explicitly investigate the role of baryons in dilepton production
in relativistic nuclear collisions. Our paper is organized as follows:
we first explicitly examine the role of the $N^*(1520)$ resonance,
which will turn out to be the largest baryon dilepton channel. We also
examine a few subtleties associated with its off-shell behaviour and its
coupling to electromagnetic radiation. We then compare rate calculations
with each other in order to get a feeling for the magnitude of the
different contributions in a somewhat idealized thermal environment. We
then model the dynamics of the full nuclear collision in the UrQMD
approach \cite{urqmd}. The transport model will cover aspects that are
related to the possible importance of pre-equilibrium dynamics, while
insuring a proper reproduction of the hadronic observables. It is
important here to insist on the following: throughout this work the
effective spectral function of the vector mesons are the vacuum ones 
{\it , i.e.} we
will work with an unmodified pion form-factor. Of course the spectral function
associated with the electromagnetic current-current correlator 
will receive contribution
from the medium, such as the Dalitz decays of mesons and baryons inside the
fireball \cite{koch_hirschegg}. Therefore  in
this language the low mass lepton pairs come from hadron reactions and Dalitz
decays.  While formally the lepton pair production
process can be linked to the imaginary part of the vector meson
self-energy \cite{gk91}, its connection to Dalitz decay is clear
\cite{wel83}. 
Bear in mind that a consistent transport treatment of in-medium spectral
functions is a topic that still requires development.  In keeping
with our focus on the role of baryons we then formulate a prediction for the
dilepton spectra measured in recent low-energy runs of the CERN SPS. We
finally conclude. 

\section{Baryonic Interactions}

Since one of the main issues we wish to address in this work is the role played
by baryons in the production of low mass lepton pairs, this section first deals 
with technical issues that arise in such calculations. The baryons will
mostly manifest themselves through their radiative decay channel into a
dilepton. To fix the ideas, we explicitly consider the case of spin 3/2
baryon resonances which will turn out to be the most important contribution
in any case.  The channel we are considering is thus 
$R\rightarrow N e^+ e^-$, where R denotes the baryon
resonance. The interaction Lagrangian is \cite{nm95}
\bea
{\cal L}_{ R N\gamma} =  \frac{ieg_1}{2M} {\bar\psi}^\mu_R \Theta(z_1)_{\mu\nu}
\gamma_\lambda\, \Gamma\, T_3\psi_N F^{\nu\lambda}   
 - \frac{eg_2}{4M^2} {\bar\psi}^{\alpha}_R\Theta_{\alpha\mu}(z_2) 
\,T_3\, \Gamma\, \partial_\nu\psi_N\, F^{\nu\mu} + h.c.\ ,
\label{rel1}
\eea
where $
\Theta_{\mu\nu}(z)= g_{\mu\nu}-1/2 (1+2z)\gamma_\mu\gamma_\nu
$. $\Gamma$ is either 1 or $\gamma_5$ depending upon the parity of the
resonances and $T$ is a $3/2 \rightarrow 1/2$ isospin 
transition operator or a $2 \times 2$ Pauli matrix,  depending on
the isospin of the resonance.
$\Psi^\mu_R$ and $\psi$ correspond to Rarita-Schwinger and
nucleon spinors respectively, $F^{\mu\nu}$ is the electromagnetic field
tensor, and $M$ is the nucleon mass.  
The influence of the off-shell parameter $z$ is seen in
calculations of electromagnetic transitions of nucleon resonances
into different multipolarities \cite{nm95}.
It should be
mentioned here that when the spin 3/2 particle is on-shell 
the results are independent of parameter $z$. Explicit  
calculation shows that the Rarita-Schwinger projection operator
\bea
\Delta_{\mu\nu}(q)=(q\!\!\!/+M_R)(-g_{\mu\nu}+\frac{2}{3}\frac{q_\mu q_\nu}
{M_R^2}+ \frac{1}{3}\gamma_\mu\gamma_\nu -\frac{1}{3}\frac{q_\mu\gamma_\nu
-q_\nu\gamma_\mu}{M_R})
\eea
which appears in the squared matrix element for the decay of the 3/2 
resonance contracted with $z$-dependent vertex terms $\gamma_\mu$
vanishes for on-shell particles. In other words, only the first term
of the $\Theta_{\mu\nu}(z)$ remains operative for vertex involving
on-shell 3/2 resonances.
The presence of the second term in the Lagrangian is important in order to
keep the electric quadrupole (E2) and magnetic dipole (M1)
transitions independent \cite{nm95}. 
Quantitatively however, we find that the contributions from 
the second term to the dilepton
yield are smaller than that from the first one by an order of magnitude.

Before going into more detailed discussions, we point out some 
differences between this work and some current popular approaches
\cite{rapp96,lenske} that rely on
a nonrelativistic (NR) reduction of 
\bea
{\cal L}=\frac{f_{RN\rho}}{m_\rho}\bar{\Psi^\mu_R}\gamma^\nu T F_{\mu\nu}\psi_N
\label{rel2}
\eea
to calculate the $\rho$ spectral function. To cast it into the form relevant
for the NR case, one neglects the lower component of the Dirac spinor and 
upon simplification one obtains

\bea
{\cal L}=\frac{f_{RN\rho}}{m_\rho}
\Psi^\dagger_R(S_k\rho_k\omega - \rho_0 S_k q_k)\psi_N
\label{nrel}
\eea
where $S_k$ ($k$ = 1, 2, 3) is the 3/2 spin transition operator\cite{weise}, 
$\rho^\mu$ = ($\rho_0$, $\rho_k$) is the $\rho$ meson field, and 
$q^\mu =(\omega, q_k)$ is its four momentum.

The approach represented by Eq. (\ref{rel1}) enables
a study of the influence of the off-shell parameter $z$. This is done later
on. By setting $z = -1/2$ and $g_2=0$ and using the 
the Vector Meson Dominance model (VMD), Eq.(\ref{rel1}) can be 
cast in the form of the Lagrangian of Eq. (\ref{rel2}). 
We note that it is satisfying to have a context where the importance of
relativistic  effects can be appreciated quantitatively. Those are found to
be of some importance in the lower invariant mass region, as we will see
shortly. 

\begin{figure}
\begin{center}
\epsfig{file=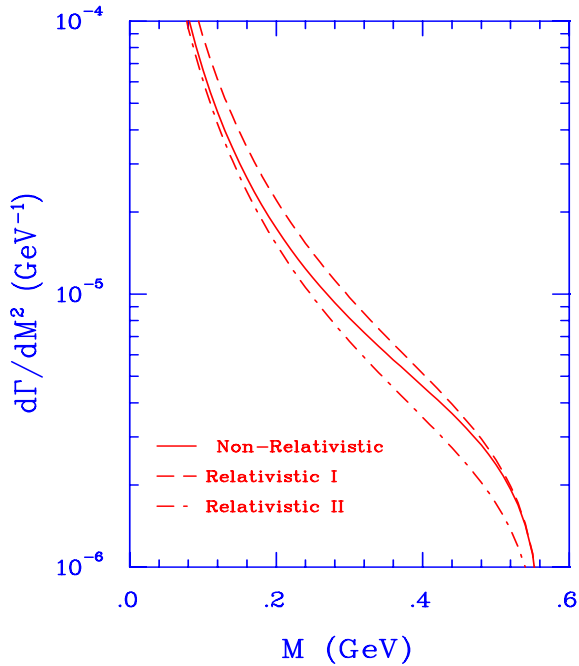,width=7.5cm,height=8.75cm,angle=0}
\end{center}
\caption{Solid and dashed lines represent results obtained with the
non-relativistic and relativistic versions of 
Eq. (\ref{rel2}). The dashed-dotted line shows 
the result obtained with Eq. (\ref{rel1}).}
\label{fig1}
\end{figure}
Let us consider the $N^\ast(1520)$ resonance. In the interaction Lagrangian of Eq.
(\ref{rel1}) we use parameters obtained in Ref. \cite{mosel}, where a fit
to pion photoproduction multipoles was performed.
To be quantitative, we take $g_1 = -1.839$, $g_2 = 0.018$, $z_1 = -0.092$
and $z_2 = -0.024$. There exists other parameter sets which yield 
comparable $\chi^2$ as far as the photoproduction data is concerned, but in
addition we will require sensible results for the calculation of radiative decay
channels.  The above quoted values of the parameter set yields the
following widths: $\Gamma_{N^\ast(1520)
\rightarrow N\gamma} = 0.78 $ MeV and $\Gamma_{N^\ast(1520)\rightarrow 
N \rho} = 22.35 $ MeV. The experimental values are 0.55$\pm$0.1 MeV and 
24.0 $\pm$ 6.0 MeV
respectively. Using VMD with the nonrelativistic
interaction Lagrangian Eq. (\ref{nrel}) and fitting its only constant in
order to reproduce $\Gamma_{N^* \rightarrow N\rho}$, one obtains $f_\rho=5.5$ 
and $\Gamma_{N \gamma}$ = 0.88 MeV. 
Using the same $f_\rho$ in the relativistic Lagrangian, Eq. (\ref{rel2}), 
produces 
$\Gamma_{N \gamma}$ = 1.13 MeV and $\Gamma_{N \rho}$ = 30 MeV.

Given a Lagrangian, the differential partial width into a lepton 
pair of invariant mass $M$ is calculated by standard techniques \cite{gale94}.
We first investigate the effects of the different interactions. We fix the
mass of the $N^*(1520)$ at its central value ({\it i.e.} at the peak of
the vacuum spectral function).  The results are shown in Fig. \ref{fig1}. 
The solid and dashed curves labeled Non-Relativistic and Relativistic I  
respectively correspond to the cases where the 
Lagrangian (\ref{rel2}) and its NR reduction
(\ref{nrel}) have been used with the same coupling constant, $f_\rho$ =
5.5. Comparing those two curves thus permits a direct assessment of 
relativistic
effects. Those are found to be largest at the photon point and 
they become smaller
at larger invariant masses, where the nucleon has a smaller
momentum. The relativistic and NR treatments should then be in agreement at
the larger invariant masses and this is indeed the case. 
The  dashed-dotted curve (Relativistic II)
represents the differential width obtained using Eq. (\ref{rel1}). The difference
between the two relativistic results appear to be a simple overall scaling.
We have verified that this overall shift in fact 
corresponds to the ratio of the
two radiative decay widths in the two models. We recall the versatility
and the success of the relativistic approach represented by 
Lagrangian (\ref{rel1}) in terms of its capability to 
reproduce pion photoproduction multipoles and decay widths into
$\gamma N$ and $\rho N$ final states. However, the quantitative
differences with a nonrelativistic treatment are overall not large. 
The relativistic approach can also lend itself
to a study of off-shell effects (owing to its $z$-dependence):  a topic
towards which we now turn.

\begin{figure}
\begin{center}
\epsfig{file=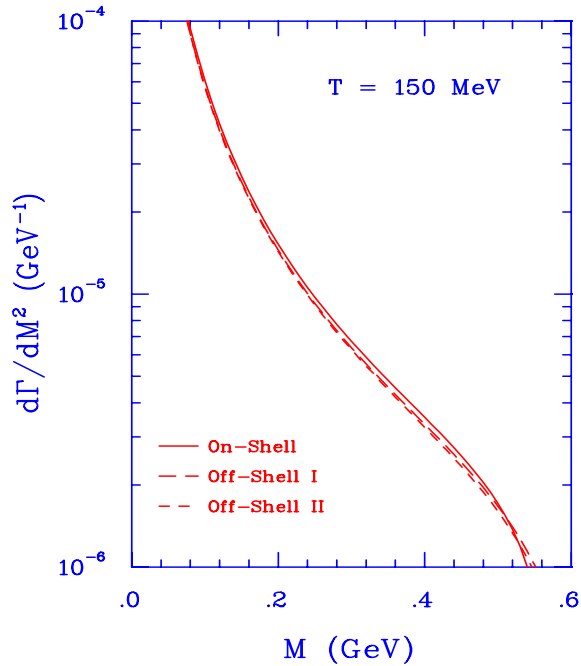,width=7.5cm,height=8.75cm,angle=0}
\end{center}
\caption{
The solid line shows results when $N^\ast(1520)$ is on-shell. The dashed-dotted
and dotted curve depicts the results where the differential width has been
integrated over $N^\ast(1520)$ spectral function with (Off-shell I)
 and with out off- shell interaction ($z = -1/2 $) respectively (Off-shell II).}
\label{fig2}
\end{figure}

We have calculated $d\Gamma/dM^2$ for $N^\ast(1520) \rightarrow N e^+ e^-$ 
in three different ways.
First using the on-shell mass (``On-Shell'', in Fig. \ref{fig2}), 
then integrating over the vacuum spectral function of this
resonant state with (``Off-Shell I''), and without (``Off-Shell II'')
the $z$-dependent interaction. This latter
case was achieved by setting $z = -1/2$. 
It is clear that off-shell
effects can arise not only from the mass distribution but also from
the associated vertex containing the factor $\Theta_{\mu\nu}(z)$. A
comparison of those three approaches is shown in Fig. \ref{fig2}. 
It is seen that off-shellness can cause differences of at most 
a few percent. Summarizing this section, one concludes that in this case
off-shell effects are not numerically important and that relativistic effects 
are slightly larger. However, a nonrelativistic treatment remains a sensible
approximation. 

\section{Rates}

Before going into the transport calculation and thereby dealing with
possible complications owing to the dynamics, we present
thermal rate calculations of various lepton pair sources and compare 
them with with each other. As discussed above, the relativistic effects
are not large although not necessarily negligible. 
Note  that the difference between 
the relativistic and non-relativistic results might become substantial, 
if evaluated in a different 
frame of reference \cite{post_mosel}. However, in the present work 
we will employ the calculated decay within a transport model, where the 
widths are always evaluated in the rest frame of the resonance and then 
simply boosted to the appropriate matter frame. Thus, in this context using 
the non-relativistic expression for the widths is a good approximation.
Using Eq. (\ref{nrel}) and VMD, the differential partial width for the decay of
baryonic resonance $R_i$ into a nucleon and a lepton pair of invariant mass
$M$ can be written as \cite{lenske}
\be
\frac{d\Gamma_{{R_i}\rho}}{dM}
=\left(\frac{f_{R_i N\rho}}{m_\rho}\right)^2\frac{1}{\pi}\,S_\Gamma\,
M\, \frac{m_N}{m_{R_i}}\,k_\rho\, A\,F_\rho (M)\,F(k_\rho^2)\ ,
\ee
where
\begin{eqnarray}
A&=&k_\rho^2\qquad {\rm for}~~ p-{\rm wave}\nonumber\\
&=&2k_\rho^2+3M^2\qquad {\rm for}~~ s-{\rm wave},
\end{eqnarray}
and $F(k_\rho^2)=\Lambda^2/(\Lambda^2+k_\rho^2)$, with
$\Lambda=1.5$ GeV. In the above, $f_{R_iN\rho}$ is the coupling
coupling constant, $S_\Gamma$ is the spin sum, and $k_\rho$ is the
momentum of the rho meson in the rest frame of the baryon
resonance. 
\begin{figure}
\begin{center}
\epsfig{file=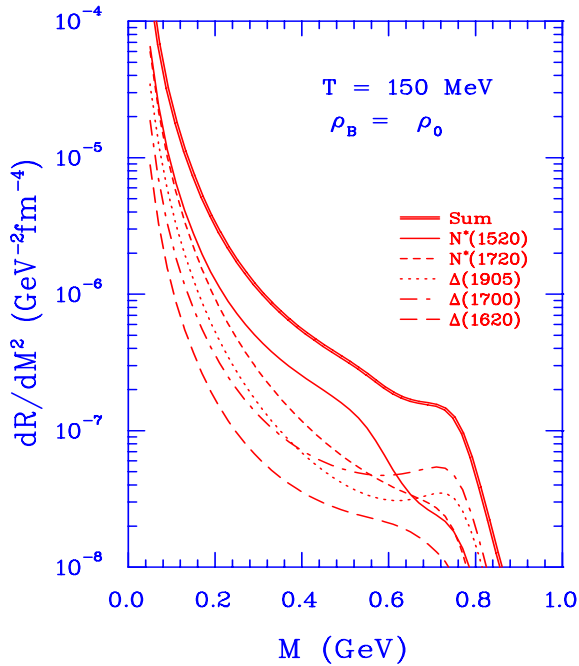,width=7.5cm,height=8.75cm,angle=0}
\end{center}
\caption{
Relative contributions of the baryonic resonances to the
dilepton production rate and their sum. }
\label{fig3}
\end{figure}
We should note that the vector dominance coupling used here for the 
$N^*(1720)$ and $\Delta(1905)$ over-estimates the measured photon decay
width \cite{friman_pirner}. Thus our results for the 
dilepton rates due to these resonances are in fact 
upper limits. 

The $\rho$ mass distribution is
\begin{equation}\label{mass}
F_\rho(M)=\frac{1}{\pi}\frac{m_\rho\Gamma_\rho(M)}{(M^2-m_\rho^2)^2
+(m_\rho\Gamma_\rho(M))^2}\ . 
\end{equation}

One uses the differential decay width to derive the differential 
dilepton production rate. For a general process $a \rightarrow b + e^+e^-$,
this is \cite{gale94}

\bea
\frac{dR_{a\rightarrow b + e^+e^-}}{dM^2}
= \frac{N m_a}{(2\pi)^2} \frac{d\Gamma_{a\rightarrow b + e^+e^-}}{dM^2}
\int_{m_a}^\infty dE_a p_a f_a(E_a)\int_{-1}^{1}dx[1 + f_b(e_b)]\ ,
\eea

where $E_b= (E_a E_b^* + p_a p_b^* x)/m_a$ and 
$E_b^*= (m_a^2 + m_b^2 -M^2)/(2 m_a)$ \cite{gale94}. In the above we 
have assumed that
$a$ and $b$ were fermions and thus the $f$'s are Fermi-Dirac distribution
functions.

\begin{figure}
\begin{center}
\epsfig{file=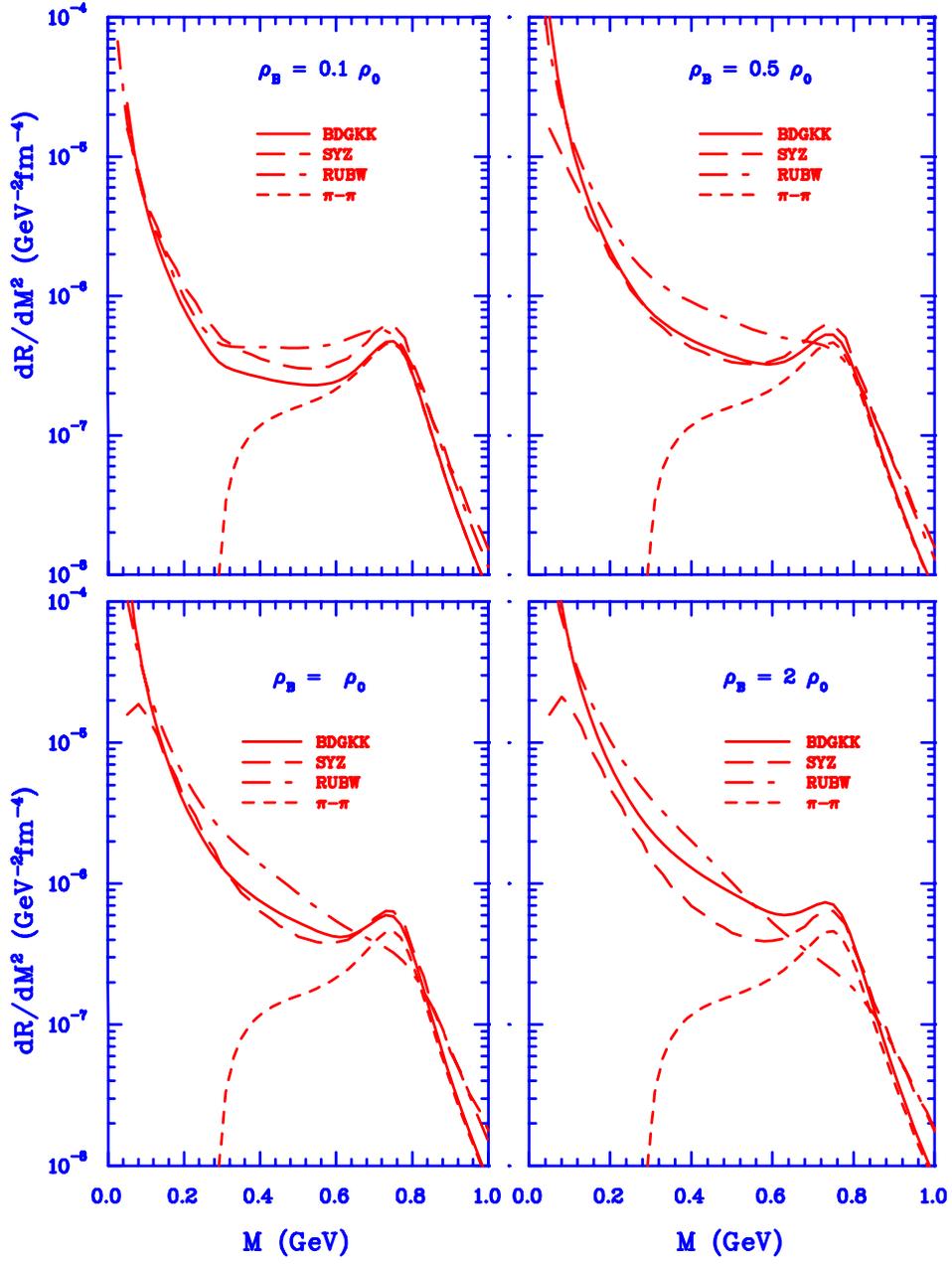,width=12.5cm,height=16.75cm,angle=0}
\end{center}
\caption{Comparison of the rates obtained in this work with that
of others at T = 150 MeV. The  short-dashed curve corresponds to $\pi$-$\pi$
annihilation only, and the full, long dashed and dashed dotted curves refer to 
the rates calculated here (DGKK), by SYZ and by RUBW respectively. 
See the main text for references.}
\label{fig4}
\end{figure}

We first calculate and compare dilepton rates associated with the 
radiative decay of
several hadronic resonances. The species we consider and the associated  rates 
are shown in Fig.~\ref{fig3}. We show representative results obtained at normal 
nuclear matter density.  Adding to the sum of baryonic contributions the
signal from  $\pi-\pi$ annihilation and that from $\omega$ decay one
obtains the net rates shown in Fig.~\ref{fig4}. 
\begin{figure}
\begin{center}
\epsfig{file=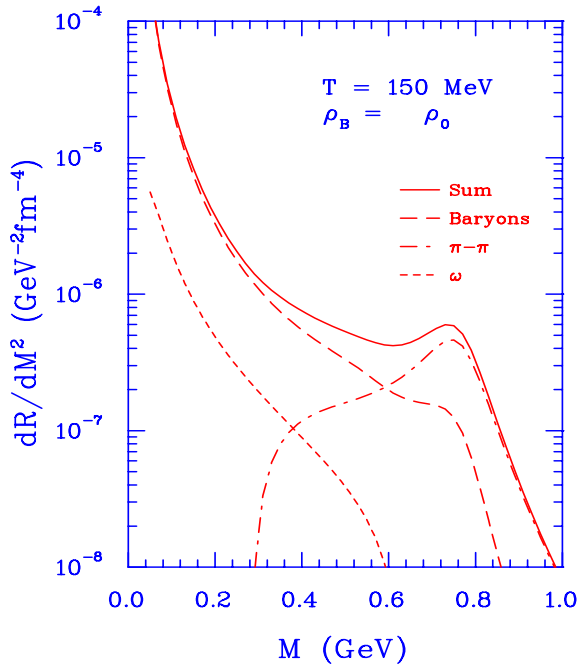,width=7.5cm,height=8.75cm,angle=0}
\end{center}
\caption{The relative importance of mesonic and
baryonic contribution to the dilepton rate is shown. The solid 
line corresponds 
to the total yield while the dashed line represents the sum over
baryonic contributions. Dashed
and dotted curves refer to the $\pi$-$\pi$ annihilation and $\omega$
Dalitz decay, respectively.}
\label{fig5}
\end{figure}
The two mesonic sources  we
have considered are actually the largest ones in the invariant mass range
we have chosen to consider. Also shown on Fig.~\ref{fig4} are the rates
obtained by Steele, Yamagishi, and Zahed (SYZ) \cite{syz} and by 
Rapp, Urban, Buballa, and Wambach (RUBW) \cite{rubw}. At very low baryonic
densities, the three rates are quite close to each other. As the density
grows our rates remain quite close to that obtained in the SYZ formalism.
At the highest baryonic density we have considered, our results
approximately lie in between those of SYZ and RUBW, for low invariant
masses. A marked difference exists in the results of RUBW, that can be
understood in terms of the 
spectral function re-summation technique leading to a suppression at 
the $\rho$ peak \cite{raga}.
To provide a reference point, the analysis of particle ratios 
for SPS-collision at 160~GeV per nucleon gives a chemical freeze-out 
temperature
of about $T= 170 \, \rm MeV$ and a baryon density 
of about $\rho = \rho_0$ (see {\it e.g.} \cite{stachel_chem}).
Finally, the relative importance of our mesonic and baryonic rates is
shown in Fig.~\ref{fig5}. The baryons dominate the low invariant mass
region while the $\pi-\pi$ channel picks up near the $\rho$ peak.
Representative results at normal nuclear matter density are shown. 
The
relative importance of mesons and baryons does not change considerably if one
uses the chemical freeze-out temperature of $170 \, \rm MeV$. It is
finally 
important to realize that the actual contribution from the $\omega$ Dalitz 
decay to the final measurements is much
higher than it appears from Fig.~\ref{fig5}: the dominant part is
due to the decay of the final state $\omega$'s.

\section{Transport calculation}

The nuclear dynamics and dilepton production is studied here within the
framework of the Ultra-Relativistic Quantum Molecular
Dynamics model, UrQMD \cite{urqmd}.
In addition to earlier molecular dynamics models 
its collision term describes the production of all
established meson and baryon resonances up to about 2~GeV with all
corresponding isospin projections  and antiparticle states. 

In the present model, light meson formation at low energies is modeled by 
multi-step processes that proceed via intermediate heavy meson
and baryon resonances and their subsequent decay \cite{urqmd}.
The pole masses, decay widths and branching ratios of all resonances 
are taken from \cite{PDG96}. However, due to the experimental uncertainties 
a certain range of parameters might be used to obtain a consistent fit
to cross section data. As an example, the production
of $\omega$ mesons is described in the UrQMD model by the formation and
the decay of the $N^\star(1900)$ resonance.  It decays in 35\% of the
cases into $N\pi$
and 55\% into $N\omega$. In line with data at SIS energies (1 AGeV), 
the $\eta$
production proceeds not only via $N^\star(1535)$, but invokes 
also nucleon resonances with masses from 1650~MeV to 2080~MeV.
A detailed description of the resonance formation cross sections and
comparisons to data is give in Ref. \cite{urqmd}.
For earlier works on dileptons within the same framework 
see {\it e.g.} \cite{ernst98a}.

Dilepton production is calculated in the following way. We consider two
distinct classes of processes. First, direct decays of vector mesons, i.e. the
$\rho$ and $\omega$ mesons. Second, the Dalitz decay of mesons and baryons
during the reaction, specifically the $\omega$ and $a_1$ meson as well as the
$N^*(1520)$. The Dalitz decay of the 
pion and eta is treated at the end of
the simulation according to Ref. \cite{song_koch}.
By treating the direct decay of the $\rho$ we implicitly also include the
contribution form the pion annihilation channel (see {\it e.g.} 
\cite{song_koch}). In order to avoid
double counting, we do not allow $\rho$ mesons created via the decay of the
$N^*(1520)$ or $a_1$-meson to decay into dileptons, since we treat the Dalitz
decay of these states explicitly. 
Since in UrQMD the $\omega$-meson does not have a $\rho-\pi$ decay channel,
there is no double counting in this channel. 
One should note at this point that
treating the Dalitz decays as a two-step process via an intermediate $\rho$
meson can lead to erroneous results in a transport description: in this case
lepton pairs below the minimum mass of $m < 2 m_\pi$ cannot be produced while
this is of course kinematically allowed for a Dalitz decay.     

In order to extract the dilepton yield from UrQMD, we extract for each of the
aforementioned channels the lifetime and four-momentum. 
The properties of the individual hadrons are given by the collision
dynamics. If a meson is produced in an elementary collision (may it be due to
string fragmentation, meson or baryon resonance decay or annihilation) it
gets a mass according to a Breit-Wigner distribution and a 4-momentum
vector according to the kinematics of this single reaction.
The meson is then assumed to travel on a straight line trajectory 
until it (i) decays or (ii) collides with another particle and its
4-momentum changes.
Any change of the 4-momentum of the meson is considered as a destruction of the
original meson and the creation of a new meson with new quantum numbers and a
new 4-momentum.
Thus for each hadron under consideration UrQMD provides a list of 
four-momenta $q$ and lifetimes $\Delta t$,  $\{ \{q,\Delta t \}_i \}_h$, where
the index $h$ labels the hadron, and $i$ labels the different four-momenta and
time-intervals for which the hadrons exists during the course of the collision.The resulting dilepton spectrum is then obtained from
\be 
\frac{d N_{l^+l^-}}{dM}= \sum_i 
\left(\Delta t_i \, \Gamma_{V\to l^+l^-}(M) \, 
\delta\left(\sqrt{q^2_i}-M\right) 
\right)_{h=V}
\ee
for the the direct decay of vector-mesons.
For the Dalitz decays, on the other hand we have
\be
\frac{d N^{\rm Dalitz}_{l^+l^-}}{dM}= \sum_i 
\left( \Delta t_i \,\frac{d\Gamma(q^2_i)_{R \to l^+ l^- X}}
{dM} \right) \ .
\ee

If absorption is negligible, this
approach is equivalent to the method of adding one dilepton (with appropriate
normalization) at each decay vertex. However, this ``shining'' method
gives better numerical statistics compared to a single decay of the vector
meson. 
To get an upper limit on the possible dilepton radiation, particles 
are allowed to couple to the dilepton channel also during their
formation time. If we insist that the particles shine only after 
their formation time is over, we have verified that the net rate 
is only reduced by $\approx$~25\%. 

\begin{figure}[hb]
\begin{center}
\epsfig{file=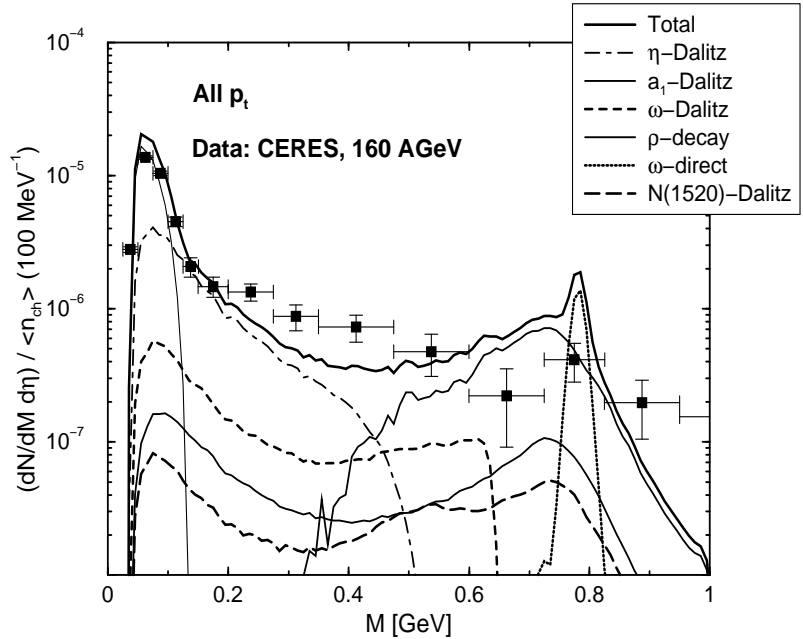,width=10.5cm,height=8.75cm,angle=0}
\caption{Invariant mass spectrum for all pair momenta, for Au-Au
collisions at 40 AGeV. Data are from 
\protect\cite{ceres96} and are shown only to set a scale. See the text for
details.}
\label{fig6}
\end{center}
\end{figure}
\begin{figure}
\begin{center}
\epsfig{file=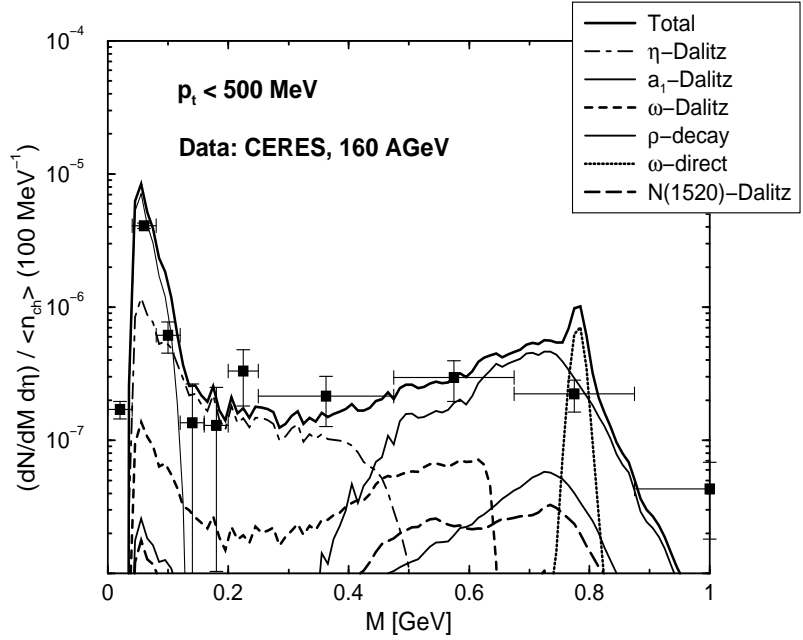,width=10.5cm,height=8.75cm,angle=0}
\caption{Same caption as for Fig. \protect\ref{fig6}, but only for low
$p_T$ lepton pairs.\label{fig7}}
\end{center}
\end{figure}
\begin{figure}
\begin{center}
\epsfig{file=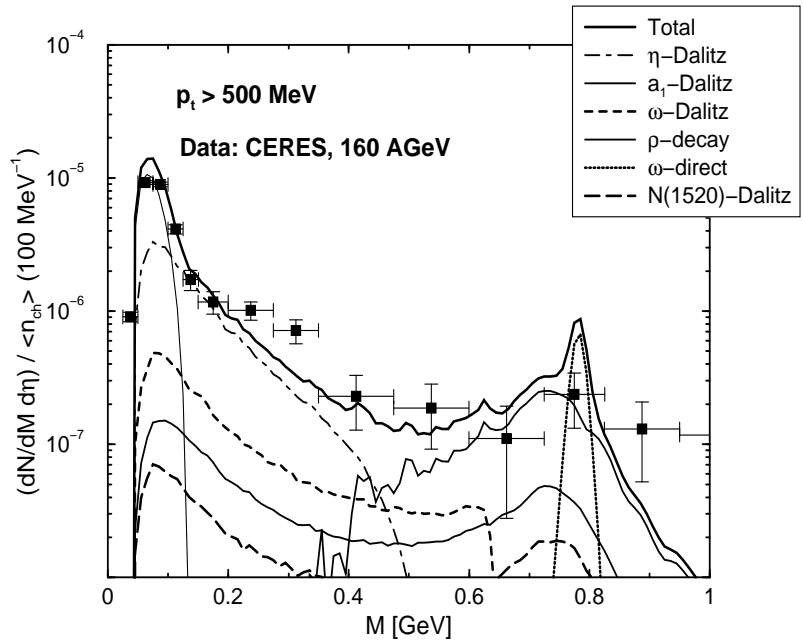,width=10.5cm,height=8.75cm,angle=0}
\caption{Same caption as for Fig. \protect\ref{fig6}, but for high 
$p_T$ lepton pairs.\label{fig8}}
\end{center}
\end{figure}

Since we are specifically concerned
with the role of the baryons in the production of lepton pair, we 
present a set of predictions for recent CERN low-energy runs where
baryon density effects should be larger than for the high energy runs. 
Those predictions  appear in Figs. \ref{fig6}, \ref{fig7} and
\ref{fig8}.

To provide a visual scale reference, we also show 
data by the CERES collaboration for 160~AGeV collisions \cite{ceres96}.
Aside from the clearly visible $\omega$-peak, we predict that there will be
little difference between the ``high-energy'' (160~AGeV) and ``low-energy''
(40~AGeV) dilepton invariant mass spectra. At an invariant mass of 400
MeV, the 160 AGeV calculation is 20\% higher (including the CERES
normalization) than the 40 AGeV one \cite{dietrich}.  
Here, we have worked with a mass
resolution of $\Delta M / M = 1 \%$. If this resolution can be achieved in
experiment, the omega-peak should be clearly visible, providing a strong
constraint for presently available model calculations.  
Furthermore, we find that the contribution of the $N^*(1520)$-Dalitz (thick
dashed curve) is very small, even at these low bombarding energy.
We finally note, that a calculation based on the model of \cite{song_koch}
with the addition of the $N^*(1520)$ channel and constrained by the  
final state hadronic spectra as predicted by  the RQMD model for a 40~AGeV
collision, gives virtually the same results. This model also is in fair
agreement with the dilepton spectra measured in the higher energy 
collisions \cite{koch_hirschegg}.

\section{Conclusion}

We see that our transport results lead to a very small baryonic
contribution to the net dilepton yield for the recently done low energy
heavy ion runs of the CERN SPS. This is to be contrasted with
the baryon influence on the net rate, as shown in Fig.~\ref{fig5}. The
reason for this difference is two-fold. First, the baryonic contribution
are severely cut down by the CERES acceptance which we apply to our
transport results in Figs.~\ref{fig6} and \ref{fig7}. Second, the baryon
content of the central rapidity region is smaller than its mesonic
content, owing to the dynamics of the stopping power systematics and of
the phase space for particle creation. The
small baryon contributions obtained here can be compared with similar
results obtained with a hydrodynamic model \cite{prakash99}. 
Therefore it really does appear that the 
ideal test to pin down the
relevant dilepton production  scenario would involve a low energy 
run. A high resolution measurement at the positions of the
vector meson vacuum masses would highlight the high baryon density. 
In such an environment the collective
behaviour of the many-body calculations leads to a
suppression of the vector meson peak \cite{raga}.  In this context we
have generated a set of predictions and the corresponding 
dilepton results of the CERN low energy runs are eagerly awaited. 
We are also looking forward to HADES measurements at the GSI
\cite{hades} where it has been shown that the dilepton contribution from 
N$^*$(1520) Dalitz decay is most important in the invariant mass 
region 0.35 $< M <$ 0.75 GeV/c$^2$ \cite{bratko}.

\acknowledgements

CG, CMK, and VK wish to acknowledge the hospitality of the Theory Institute
of the University of Giessen, where this work was started. We also would like
to thank P. Huovinen for furnishing the SYZ and RUBW rates used in
Fig.~\ref{fig4}. 
The work of CG and AKDM is supported in part 
by the Natural Sciences and Engineering Research Council of Canada
and in part the Fonds FCAR of the Qu\'ebec Government. 
The work of CMK is supported by the National Science Foundation 
under Grant No. PHY-9870038, the Welch Foundation under Grant No. A-1358, 
the Texas Advanced Research Program under Grants No. FY97-010366-0068 and
FY99-010366-0081, and the Alexander von Humboldt Foundation. 
VK and MB are supported by the Director, Office of
Energy Research, division of Nuclear Physics of the Office of High
Energy and Nuclear Physics of the U.S. Department of Energy under
Contract No. DE-AC03-76SF00098. MB is supported by a Feodor-Lynen 
fellowship of the Alexander von Humboldt Foundation.

\end{document}